# Polymorphism and Magnetism in a Kitaev Honeycomb Cobaltate KCoAsO$_4$

Yuya Haraguchi[1,*], Daisuke-Nishio Hamane[2], and Hiroko Aruga Katori[1]

[1]*Department of Applied Physics and Chemical Engineering, Tokyo University of Agriculture and Technology, Koganei, Tokyo 184-8588, Japan*

[2]*The Institute for Solid State Physics, The University of Tokyo, Kashiwa, Chiba 277-8581, Japan*

Corresponding author: *chiyuya3@go.tuat.ac.jp

**Abstract**

We report the synthesis, crystal structure, and magnetic properties of a new Kitaev honeycomb cobaltate, KCoAsO$_4$, which crystallizes in two distinct forms: $P2/c$ and $R\bar{3}$ space groups. Magnetic measurements reveal ordering temperatures of ~14 K for the $P2/c$ structure and ~10.5 K for the $R\bar{3}$ structure. The $P2/c$-type KCoAsO$_4$ sample exhibits a complex temperature-field phase diagram, including a field-induced phase, while the $R\bar{3}$-type KCoAsO$_4$ shows a simpler phase diagram with a single magnetically ordered phase. The observed differences in magnetic properties are attributed to subtle structural variations, strongly suggesting that local structural changes play a crucial role in determining the magnetism of cobaltate-based Kitaev materials.

**Introduction**

Spin frustration serves as a cornerstone in the study of unconventional magnetic ground states within condensed matter physics. Geometrical frustration can emerge naturally in specific lattice geometries, particularly in non-bipartite structures like triangular and kagome lattices. These geometries inherently prevent Néel-type antiferromagnetic order, fostering the formation of unconventional quantum states, such as quantum spin liquids (QSLs), which are defined by long-range quantum entanglement rather than classical magnetic order [1-7].

Recently, honeycomb lattices have become central to QSL research. Despite being bipartite, these lattices introduce a unique form of spin frustration through bond-dependent anisotropic interactions, often referred to as Kitaev interactions [8-11]. Remarkably, in such Kitaev systems, an exactly solvable QSL phase emerges, characterized by fractional excitations: itinerant Majorana fermions and localized Z$_2$ fluxes, which serve as the elementary excitations of the ground state [8-11]. Kitaev interactions are theoretically predicted to occur between magnetic ions with spin-orbit-entangled $J_{\text{eff}}$ = 1/2 pseudospins [9]. In edge-sharing metal-ligand octahedra, quantum interference between multiple superexchange pathways can cancel out the typical isotropic ferromagnetic/antiferromagnetic Heisenberg interaction, leaving behind the Kitaev interaction. This mechanism has been proposed for 4d ruthenates and 5d iridates with d$^5$ electron configurations, which have been the focus of extensive experimental efforts including the development of novel materials [10-11].

More recently, despite the weak spin-orbit coupling of $3d^7$ ions like $Co^{2+}$, they have been recognized as potential hosts for Kitaev physics due to the presence of a $J_{eff} = 1/2$ Kramers doublet in the ground state [12-16]. A detailed theoretical analysis of the superexchange interactions reveals that ferromagnetic and antiferromagnetic Heisenberg interactions along the different superexchange pathway effectively cancel each other, making the Kitaev interaction dominant and thereby realizing the Kitaev-Heisenberg model [12-16]. Moreover, the ratio of Kitaev-Heisenberg interaction parameters can be tuned by Hund's coupling and crystal field splitting, suggesting that realistic crystal structures may host a QSL [12,13,16]. This prediction has prompted a re-examination of honeycomb cobaltates, such as $Na_2Co_2TeO_6$ [17-21], $Na_3Co_2SbO_6$ [17,21,22], $Ca_3Co_2SbO_6$ [23,24], $BaCo_2(AsO_4)_2$ [25-28], and $BaCo_2(PO_4)_2$ [29,30], within the framework of the Kitaev-Heisenberg model. These insights have also spurred the development of new materials, such as $Li_3Co_2SbO_6$ [31] and $CaCo_2TeO_6$ [32], leading to a surge of interest in investigating Kitaev physics in cobaltates.

However, recent theoretical and experimental studies have raised doubts about whether cobaltates can truly be regarded as Kitaev candidates. Theoretical investigations of $BaCo_2(AsO_4)_2$ have shown that its spin model is more accurately described by an XXZ model, where the third-nearest-neighbor Heisenberg interaction $J_3$ on a honeycomb lattice plays a significant role, and the Kitaev interaction is negligible [33]. Neutron inelastic scattering measurements further support this finding, showing that the Kitaev interaction is nearly absent and that the system is well explained by the $XXZ$-$J_1$-$J_3$ model on honeycomb lattice [28]. Additionally, the 1/3 magnetization plateau observed during isothermal magnetization processes at cryogenic temperatures is successfully reproduced by the $XXZ$-$J_1$-$J_3$ honeycomb model [28]. Conversely, ab initio calculations have shown that in $Na_3Co_2SbO_6$ and $Na_2Co_2TeO_6$, the Kitaev interaction becomes comparable in strength to the Heisenberg interaction [33]. This theoretical insight is further supported by inelastic neutron scattering, which demonstrates that in both compounds, the ferromagnetic Kitaev interaction is dominant, accompanied by a comparable antiferromagnetic Heisenberg interaction [34]. Despite the structural similarities shared by honeycomb cobaltate systems, their spin models exhibit differences that stem from direct Co-Co exchange interactions—an aspect that was initially overlooked in the Kitaev mechanism for cobaltates [14]. Furthermore, the weakening of superexchange interactions that give rise to Kitaev interactions has been theoretically demonstrated to be influenced by trigonal crystal field [33], and this is now being increasingly supported by experimental evidence [24]. Thus, resolving the ongoing debate over the relevance of the Kitaev model to honeycomb cobaltates will require studies on a broader range of materials and the development of new candidate systems to further advance our understanding of Kitaev physics.

In this letter, we report the synthesis, crystal structure, and magnetic properties of the novel cobaltate-based Kitaev candidate material, KCoAsO$_4$. The crystal structure crystallizes in two distinct forms: $P2/c$ and $R\bar{3}$ space groups, distinguished by the presence or absence of distortions in the honeycomb network. At low magnetic fields, $P2/c$-type KCoAsO$_4$ shows magnetic ordering around 14 K, while $R\bar{3}$-type KCoAsO$_4$ shows magnetic order at 10.5K. Isothermal magnetization measurements at 2 K reveal two magnetic anomalies in $P2/c$-type KCoAsO$_4$ before reaching saturation, whereas $R\bar{3}$-type KCoAsO$_4$ shows a single anomaly at saturation. The $R\bar{3}$-type KCoAsO$_4$ presents a straightforward temperature-field phase diagram with one magnetically ordered phase, while $P2/c$-type KCoAsO$_4$ displays a more complex diagram, including a magnetic-field-induced phase. The observed differences in the magnetic properties between the two structures are likely attributed to subtle structural variations, strongly supporting the notion that slight changes in local structure play a critical role in determining the magnetism of cobaltate-based Kitaev materials.

**Experimental methods**

Monoclinic KCoAsO$_4$ was synthesized using a standard hydrothermal method. The starting materials were As$_2$O$_5 \cdot x$H$_2$O (>99.99%, Sigma-Aldrich), Co(OH)$_2$ (99.9%, FUJIFILM Wako Pure Chemical Corporation), KOH (99%, FUJIFILM Wako Pure Chemical Corporation), and deionized water. In a 30 mL PTFE beaker, 200 mg of Co(OH)$_2$, 430 mg of KOH, 600 mg of As$_2$O$_5 \cdot x$H$_2$O, and 3 mL of deionized water were combined. The beaker was heated at 230°C for 60 hours in a stainless-steel hydrothermal autoclave. After the reaction, the resulting sample was washed several times with deionized water, followed by washes with acetone twice, and then dried at room temperature to obtain a pink powder. Trigonal KCoAsO$_4$ was synthesized by heating the monoclinic KCoAsO$_4$ powder in air at 300°C for 12 hours. These products were characterized by powder x-ray diffraction (XRD) experiments in a diffractometer with Cu-K$\alpha$ radiation, and chemical analysis was conducted using a scanning electron microscope (JEOL IT100) equipped with an energy dispersive x-ray spectroscope (EDX) with 15 kV, 0.8 nA, 1 μm beam diameter). In addition, the cell parameters and crystal structure were refined by the Rietveld method using the Z-RIETVELD software [35]. The temperature dependence of the magnetization was measured under several magnetic fields using the magnetic property measurement system (MPMS; Quantum Design) at the Institute for Solid State Physics, the University of Tokyo.

**Results and Discussion**

The room-temperature powder x-ray diffraction patterns of hydrothermally synthesized KCoAsO$_4$ and the sample after thermal treatment at 300°C are shown in Figs. 1(a) and (b), respectively. The peaks are consistently indexed to reflections corresponding to the $P2/c$ space group with monoclinic lattice parameters $a = 8.74355(8)$ Å, $b = 5.04192(5)$ Å, $c = 20.3188(1)$ Å, and $\beta = 108.5760(8)°$, and to the $R\bar{3}$ space group with trigonal lattice parameters $a = 5.039567(9)$ Å, and $c = 28.76421(5)$ Å. The chemical composition determined by EDX is K:Co:As = 0.98(2):1.04(1):0.98(1) for monoclinic $KCoAsO_4$, and K:Co:As = 0.96(2):1.02(2):1.00(1) for trigonal $KCoAsO_4$, which are close to the ideal stoichiometry. These results confirm the successful synthesis of polymorphic $KCoAsO_4$. Moreover, our Rietveld refinements for the $P2/c$ and $R\bar{3}$ structures show excellent convergence as shown in Figs. 1(a) and (b), with the refined crystallographic parameters listed in Tables I and II.

The $R\bar{3}$-type $KCoAsO_4$ has the same structure as the previously reported $KNiAsO_4$ [36,37] and features a similar layered arrangement to the Kitaev candidate material $BaCo_2(AsO_4)_2$ [25]. Notably, crystallization in the $P2/c$ structure has not been reported for either $KNiAsO_4$ [36,37] or $BaCo_2(AsO_4)_2$ [25], suggesting that the $P2/c$ phase obtained via hydrothermal synthesis is a metastable phase formed through kinetically controlled reactions. This conclusion is further supported by the irreversible transformation from the $P2/c$ structure to the $R\bar{3}$ structure upon annealing at a relatively low temperature of 300°C. Additional evidence for the greater stability of the $R\bar{3}$ structure compared to the $P2/c$ structure was obtained through first-principles structural optimization using the QUANTUM ESPRESSO package [38]. Even when monoclinic distortion was introduced into the initial structure, structural optimization consistently led to the relaxation of the monoclinic distortion, converging to the $R\bar{3}$ structure. This result further confirms the thermodynamic stability of the $R\bar{3}$ structure.

The $P2/c$ structure exhibits significant distortion. As shown in the upper panel of Fig. 2, the Co honeycomb network is split into two distinct Co sites (Co1 and Co2) due to monoclinic distortion, resulting in four different Co-Co bond lengths. In contrast, as shown in the bottom panel of Fig. 2, the $R\bar{3}$ structure, which lacks monoclinic distortion, has only one Co-Co bond length of 2.916 Å, forming an undistorted, ideal honeycomb lattice. The distortions in the $CoO_6$ octahedra for both structures can be quantified using the parameters of quadratic elongation $\lambda$ [39],

$$\langle \lambda \rangle = \sum_{i=1}^{6} \frac{l_i/l_0}{n}, \tag{1}$$

where $n$ is the coordination number of anions around the central cation, $l_i$ is the bond length between the central cation and the $i$-th coordinating anions, and $l_0$ is the bond length in a polyhedron with $O_h$ symmetry, whose volume is equal to that of the distorted polyhedron, and bond angle variance $\sigma^2$ [39],

$$\sigma^2 = \sum_{i=1}^{m} \frac{(\varphi_i - \varphi_0)^2}{m-1}, \tag{2}$$

where $m$ is the number of anion-cation-anion bond angle, with $m = 12$ for octahedra, $\varphi_i$ is the $i$th bond angle of the distorted coordination polyhedron, and $\varphi_0$ is the ideal bond angle in a polyhedron with $O_h$ symmetry; for octahedra, $\varphi_0 = 90$ deg. In the $P2/c$ structure, the Co1 site has $\lambda = 1.0257$ and $\sigma^2 = 80.381$ deg$^2$, while the Co2 site has $\lambda = 1.0236$ and $\sigma^2 = 63.747$ deg$^2$.

In contrast, the Co site in the $R\bar{3}$ structure has λ = 1.0157 and σ² = 47.752 deg². These results demonstrate that, compared to the $R\bar{3}$ structure, the $P2/c$ structure is significantly more distorted, both in the global honeycomb network and in the local $CoO_6$ octahedral structure.

The temperature dependence of the inverse magnetic susceptibility (1/χ) for powder samples of $P2/c$-type $KCoAsO_4$ and $R\bar{3}$-type $KCoAsO_4$ is shown in Fig. 3(a). At high temperatures, the 1/χ data exhibit a linear relationship with $T$. Curie-Weiss fitting in the range of 200 to 300 K yields an effective paramagnetic moment $\mu_{eff}$ = 5.079(3) $\mu_B$ and a Weiss temperature $\theta_W$ = 46.6(3) K for $P2/c$-type $KCoAsO_4$, and $\mu_{eff}$ = 5.137(10) $\mu_B$ with $\theta_W$ = 35.5(8) K for $R\bar{3}$-type $KCoAsO_4$. The estimated $\mu_{eff}$-values for both compounds align well with the spin-orbit-entangled value expected for $Co^{2+}$ with $J_{eff}$ = 1/2 [11]. Notably, the positive values of $\theta_W$ for both compounds indicate dominant ferromagnetic coupling. The primary origin of these ferromagnetic interactions is likely Kitaev-type bond-anisotropic interactions or direct Heisenberg isotropic interactions between Co ions.

As shown in Fig. 3(b), at low temperatures, the χ data exhibit a sharp decrease, signaling antiferromagnetic ordering. The Néel temperatures $T_N$ are estimated to be 14 K and 10.5 K for $P2/c$-type $KCoAsO_4$ and $R\bar{3}$-type $KCoAsO_4$, respectively, based on the peak of d(χ$T$)/d$T$ [see Figs. 4(b) and (f)], which is commonly referred to as the Fisher heat capacity, reflecting the relationship between the magnetic heat capacity and magnetic susceptibility, as $C_p(T)$ ~ d(χ$T$)/d$T$ [39].

To further clarify the nature of these magnetic orders, we measured the temperature dependence of magnetic susceptibility under various magnetic fields and the isothermal magnetization process at different temperatures for both samples. In Fig. 4(a), the magnetic susceptibility of $P2/c$-type $KCoAsO_4$ is shown under various applied magnetic fields. The data reveal that the temperature of the magnetic anomaly, indicative of a transition, shifts to lower values as the applied field increases, strongly suggesting the presence of an antiferromagnetic transition in the material. Similarly, in Fig. 4(b), the peaks in the d(χ$T$)/d$T$ data in $P2/c$-type $KCoAsO_4$ shift to lower temperatures as the magnetic field increases. Interestingly, the d(χ$T$)/d$T$ data for 2.5 T and 2.75 T exhibit a shoulder-like structure at lower temperatures relative to the main peak, suggesting the presence of multi-step magnetic anomalies in this field range.

The isothermal magnetization process in $P2/c$-type $KCoAsO_4$, depicted in Fig. 4(c), also captures these multi-step magnetic anomalies. At 2 K, the magnetization shows a sharp increase around 2.5 T, followed by a more gradual rise near 3 T, and then another sharp increase at approximately 3.6 T. These features are further highlighted in the d$M$/d$H$ plot shown in Fig. 4(d), where the 2 K data clearly exhibit a double peak. As the temperature increases, these double peaks shift toward lower fields, and by 12 K, they merge to the point of being indistinguishable. By 14 K, the peaks disappear entirely, indicating the suppression of the field-induced phase transition. In the high-field region, the magnetization gradually increases due to the influence of van Vleck paramagnetism. By accounting for the van Vleck contribution and performing a linear fit in this region, then extrapolating to 0 T, the intrinsic magnetization is estimated to be approximately 2.25 $\mu_B$. Since these values align with the expected magnetic moment for $Co^{2+}$ ions with a $J_{eff} = 1/2$ pseudospin [41-43], it can be concluded that $J_{eff} = 1/2$ pseudospins with g ~ 4.5 is fully polarized. Thus, the high-field step in the two-step field-induced phase transition is identified as a transition from a magnetically ordered phase to a forced ferromagnetic phase.

In contrast, the magnetic transition in $R\bar{3}$-type $KCoAsO_4$ is simpler compared to that of $P2/c$-type $KCoAsO_4$. As illustrated in Fig. 4(e), similar to the $P2/c$ sample, the temperature at which a sharp decrease in magnetization—indicative of antiferromagnetic ordering—occurs shifts to lower temperatures as the applied magnetic field increases, with the magnitude of the decrease also becoming less pronounced. In Fig. 4(f), no evidence of a two-step magnetic anomaly, as seen in $P2/c$-type $KCoAsO_4$, is observed. Instead, the peak structure remains broad, maintaining a single peak up to 2.5 T, where the magnetic ordering is suppressed.

In the isothermal magnetization process of $R\bar{3}$-type $KCoAsO_4$ at 2 K, shown in Fig. 4(g), magnetization sharply increases around 2.3 T, followed by a gradual rise. The peak in the d$M$/d$H$ data shifts to lower temperatures as the measurement temperature increases, shown in Fig. 4(h). Although a shoulder-like feature in the d$M$/d$H$ data is observed at slightly higher magnetic fields than the main peak at 9 K and 10 K, no distinct double-peak structure, like the one seen in Fig. 4(d), is present, and there is no clear evidence supporting the existence of a magnetic field-induced phase transition. Similar to $P2/c$-type $KCoAsO_4$, by accounting for the van Vleck contribution in $R\bar{3}$-type $KCoAsO_4$, the magnetization in the higher magnetic field region is estimated to be approximately 2.2 $\mu_B$. This value indicates that the $J_{eff} = 1/2$ pseudospin of $Co^{2+}$ with g ~ 4.4 is fully polarized [41-43], further confirming that the field-induced phase transition is from a magnetically ordered phase to a forced ferromagnetic phase.

Figures 5(a) and (b) present the temperature versus magnetic field phase diagrams for both the $P2/c$ and $R\bar{3}$ samples. In $P2/c$-type $KCoAsO_4$, a clear magnetic field-induced phase is observed, whereas $R\bar{3}$-type $KCoAsO_4$ shows only a single magnetically ordered phase in its phase diagram. Additionally, the magnetic ordering temperature and the suppression temperature of the magnetic transition in $R\bar{3}$-type $KCoAsO_4$ are significantly lower than those in $P2/c$-type $KCoAsO_4$.

Here, we discuss the relationship between the structure and magnetic properties. As noted above, the Co honeycomb network in $P2/c$-type $KCoAsO_4$ consists of four distinct Co-Co bonds, each with a different balance of Kitaev-Heisenberg interactions, resulting in a highly complex spin model. In addition to the oxygen-mediated superexchange interaction, direct Heisenberg exchange arising from Co-Co orbital overlap plays a significant role, driving the system away from the Kitaev limit and stabilizing magnetic order [14]. The Co1-Co1 and Co2-Co2 bond distances are relatively long, while the two types of Co1-Co2 bonds are much shorter, suggesting that the direct Heisenberg exchange along these shorter bonds may dominate the spin interactions. The observed magnetic field-induced phase transition is likely driven by the complex spin model, rather than by a simple Kitaev-Heisenberg model. Regarding multi-step magnetic anomalies, a similar three-step transition has been observed in Na$_2$Co$_2$TeO$_6$, where two Co sites are present. Given that $P2/c$-type KCoAsO$_4$ also contains two distinct Co sites, we strongly expect that this study will stimulate future theoretical developments to clarify how the number of magnetic sites influences the Kitaev-Heisenberg model.

In $R\bar{3}$-type KCoAsO$_4$, the suppression of both the magnetic ordering temperature and the field-induced transition field, compared to $P2/c$-type KCoAsO$_4$, may be due to the reduced distortion of the CoO$_6$ octahedra, which brings the system closer to a quantum critical point near a spin-disordered state. For Co ions in octahedral coordination, the trigonal crystal field splits the triply degenerate $t_{2g}$ orbitals into doubly degenerate $e_g^\pi$ and nondegenerate $a_{1g}$ orbitals. The $t_{2g}$-splitting weakens the $t_{2g}$-$p$-$e_g$ superexchange transferring process, which is a key mechanism for Kitaev interactions in cobaltates, thereby leading to a relative increase in non-Kitaev terms [33]. As previously mentioned, the CoO$_6$ octahedra in $R\bar{3}$-type KCoAsO$_4$ are less distorted than those in $P2/c$-type KCoAsO$_4$, indicating smaller trigonal distortion. These findings suggest that both global and local structural differences play a significant role in shaping the Kitaev physics of cobaltates, as evidenced by the relationship between magnetic properties and crystal structure in these polymorphic materials.

To gain a deeper understanding of the Kitaev physics in the polymorphic KCoAsO$_4$ series, establishing a reliable method for growing single crystals is essential. Currently, we have successfully grown microcrystals weighing less than 10 μg, but this is below the detection limit for magnetization measurements using MPMS. With the development of techniques for growing larger crystals, however, more detailed investigations into their magnetic properties will become possible.

**Summary**

We present a detailed study on KCoAsO$_4$, a novel cobalt-based Kitaev candidate, examining its synthesis, crystal structure, and magnetic properties. KCoAsO$_4$ crystallizes in two space groups: $P2/c$ and $R\bar{3}$. The $P2/c$ structure exhibits notable distortions, leading to two Co sites and four distinct Co-Co bond lengths, while the $R\bar{3}$ structure forms an ideal honeycomb lattice. These structural variations significantly affect their magnetic properties: $P2/c$-type KCoAsO$_4$ shows a complex spin model with multi-step magnetic anomalies and field-induced phase transitions, while $R\bar{3}$-type KCoAsO$_4$ exhibits simpler magnetic behavior. In the $R\bar{3}$ structure, reduced CoO$_6$ octahedral distortion lowers the magnetic ordering temperature and suppresses field-induced transitions, bringing the system closer to a quantum critical point than the $P2/c$ structure. Our findings highlight the importance of both global and local structural differences in determining the magnetic behavior of honeycomb cobaltates, particularly in the context of

Kitaev physics.

**Acknowledgments**

This work was supported by JST PRESTO Grant Number JPMJPR23Q8 (Creation of Future Materials by Expanding Materials Exploration Space) and JSPS KAKENHI Grant Numbers. JP23H04616 (Transformative Research Areas (A) "Supra-ceramics"), JP24H01613 (Transformative Research Areas (A) "1000-Tesla Chemical Catastrophe"), JP22K14002 (Young Scientific Research), and JP24K06953 (Scientific Research (C)). Part of this work was carried out by joint research in the Institute for Solid State Physics, the University of Tokyo (Project Numbers 202306-GNBXX-0128, 202311-GNBXX-0018, 202306-MCBXG-0094 and 202306-MCBXG-0070).

**Data availability**

The datasets generated and analyzed during the current study are available from the corresponding author.

**TABLE I** Crystallographic parameters for $P2/c$-type KCoAsO$_4$ determined from powder x-ray diffraction experiments. The obtained monoclinic lattice parameters are $a$ = 8.74355(8) Å, $b$ = 5.04192(5) Å, $c$ = 20.3188(1) Å, and $\beta$ = 108.5760(8)°. $B$ is the atomic displacement parameter. $g$ is the occupancy factor.

| atom | site | $x$ | $y$ | $z$ | $B$ (Å$^2$) |
| --- | --- | --- | --- | --- | --- |
| K1  | 4$g$ | 0.1946(7)  | 0.016(2)    | 0.8130(2)    | 1.14(16) |
| K2  | 4$g$ | 0.3036(7)  | 0.538(2)    | 0.1795(2)    | 1.15(16) |
| Co1 | 4$g$ | 0.1595(5)  | 0.5016(18)  | 0.99069(16)  | 0.86(11) |
| Co2 | 4$g$ | 0.3255(5)  | -0.0011(2)  | 0.49696(17)  | 0.79(11) |
| As1 | 4$g$ | 0.0658(5)  | -0.0017(8)  | 0.58870(12)  | 0.68(6)  |
| As2 | 4$g$ | 0.4374(5)  | 0.5010(13)  | 0.4104(15)   | 1.14(8)  |
| O1  | 4$g$ | 0.0535(17) | 0.339(3)    | 0.5671(5)    | 0.40(12) |
| O2  | 4$g$ | 0.1384(18) | 0.041(4)    | 0.6780(4)    | 0.50(11) |
| O3  | 4$g$ | 0.1410(17) | 0.188(4)    | 0.4267(5)    | 1.0(3)   |
| O4  | 4$g$ | 0.2117(16) | 0.133(3)    | 0.0587(4)    | 0.40(13) |
| O5  | 4$g$ | 0.259(4)   | 0.255(8)    | 0.9161(9)    | 0.71(10) |
| O6  | 4$g$ | 0.3607(16) | 0.371(3)    | 0.5658(4)    | 0.5(2)   |
| O7  | 4$g$ | 0.3909(19) | 0.445(4)    | 0.3313(4)    | 0.42(13) |
| O8  | 4$g$ | 0.470(2)   | 0.146(4)    | 0.4382(5)    | 1.1(2)   |

**TABLE II** Crystallographic parameters for $R\bar{3}$-type KCoAsO$_4$ determined from powder x-ray diffraction experiments. The obtained trigonal lattice parameters are $a$ = 5.039567(9) Å, and $c$ = 28.76421(5) Å. $B$ is the atomic displacement parameter. $g$ is the occupancy factor.

| atom | site | $x$ | $y$ | $z$ | $B$ (Å$^2$) |
| --- | --- | --- | --- | --- | --- |
| K  | 6$c$  | 0          | 0         | 0.29014(2)   | 0.86(3)   |
| Co | 6$c$  | 0          | 0         | 0.163354(19) | 1.08(18)  |
| As | 6$c$  | 0          | 0         | 0.440717(14) | 0.984(14) |
| O1 | 18$f$ | -0.0117(5) | 0.6820(4) | 0.11978(2)   | 0.29(4)   |
| O2 | 6$c$  | 0          | 0         | 0.37859(5)   | 0.51(6)   |

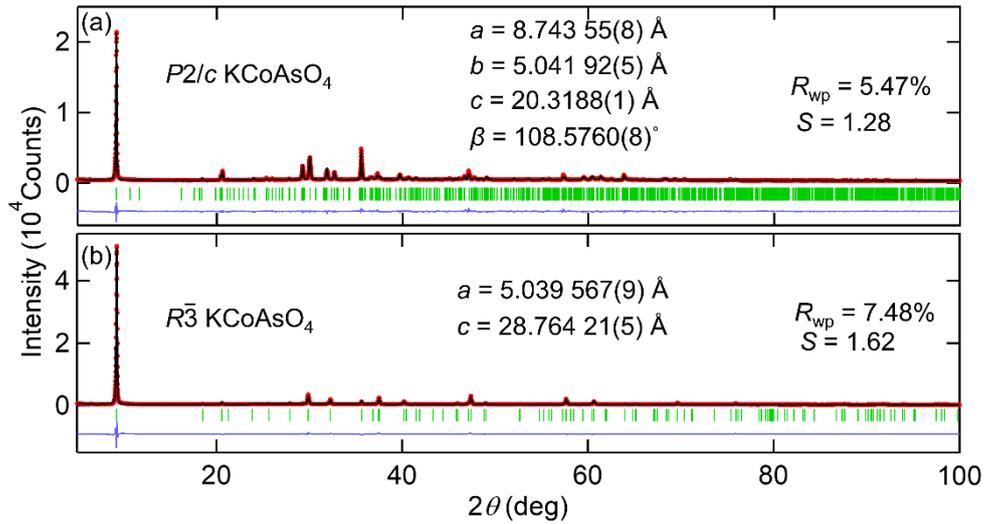

**Fig. 1** The result of Rietveld refinement of (a) $P2/c$-type KCoAsO4 and (b) $R\bar{3}$-type KCoAsO4. The observed intensities (red circles), calculated intensities (black line), and their differences (blue curve at the bottom) are shown. The green vertical bars indicate the positions of Bragg reflections, respectively.

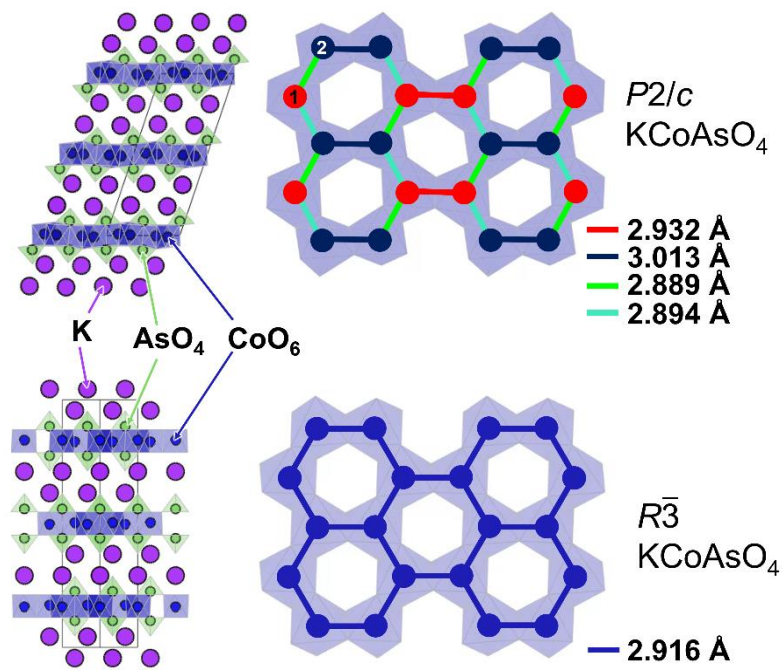

**Fig. 2** The crystal structure of *P*2/*c*-type KCoAsO₄ and *R*3̄-type KCoAsO₄ viewed along the *c* axis and the honeycomb layer viewed along the honeycomb plane. The colors and numbers assigned to the Co ions in the *P*2/*c* structure distinguish the different crystallographic sites, where four Ir-Ir bonds of varying lengths are depicted distinctly.

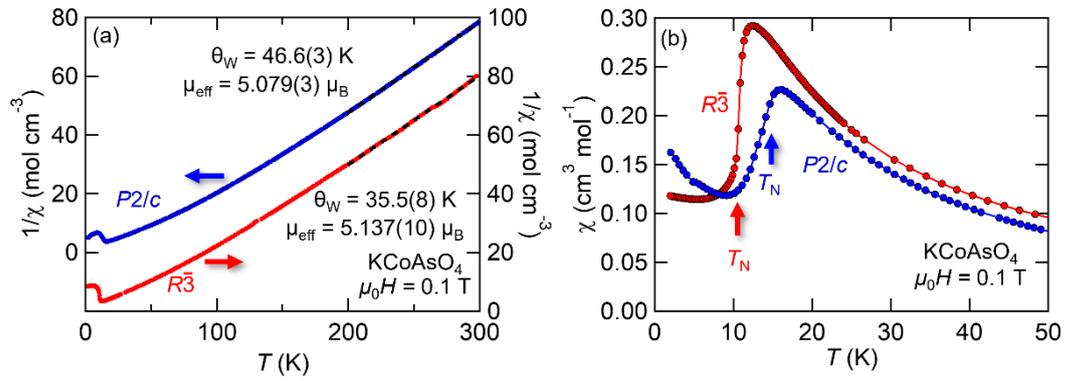

**Fig. 3** (a) Temperature dependence of inverse magnetic susceptibility $1/\chi$ for $P2/c$-type KCoAsO$_4$ and $R\bar{3}$-type KCoAsO$_4$. The dotted line represents a fit to the Curie-Weiss model. (b) Temperature dependence of the magnetic susceptibility $\chi$ in the low temperature region. Arrows indicates the magnetic ordering temperature, determined by the peak of $d(\chi T)/dT$.

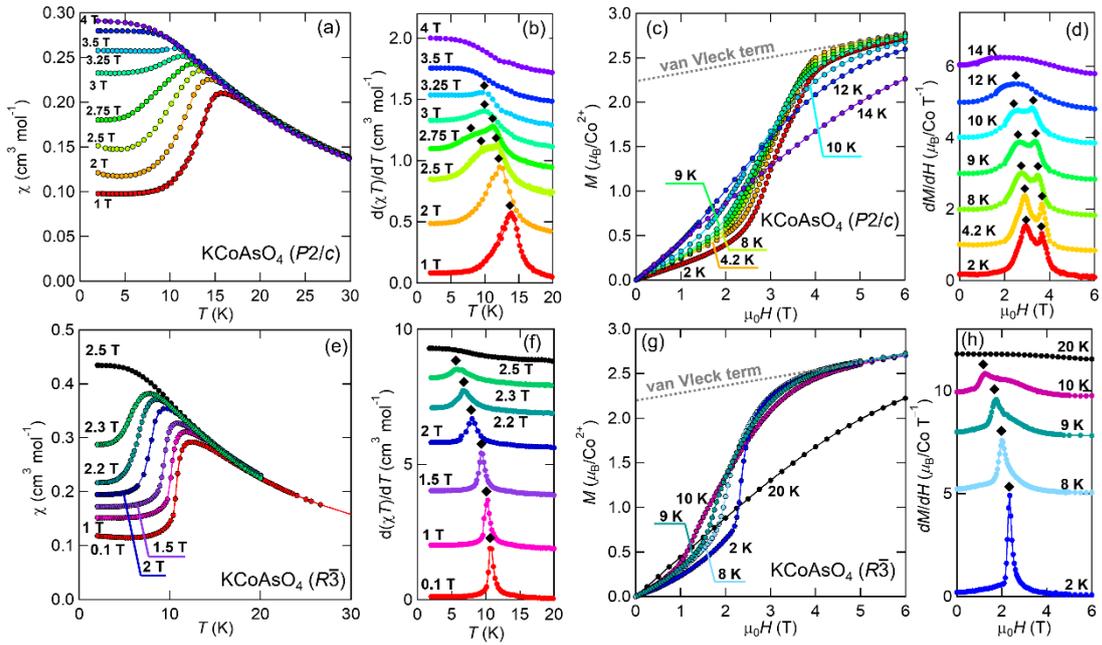

**Fig. 4** Comparison of magnetization for for $P2/c$-type KCoAsO$_4$ and $R\bar{3}$-type KCoAsO$_4$. (a) Temperature derivative of χ data and (b) d($\chi T$)/d$T$ data measured under various magnetic fields, (c) isothermal magnetization $M$ data, and (d) d$M$/d$H$ data for $P2/c$-type KCoAsO$_4$. Similarly, (e) χ-data and (f) d($\chi T$)/d$T$-data, (g) $M$-data, and (h) d$M$/d$H$-data for $R\bar{3}$-type KCoAsO$_4$

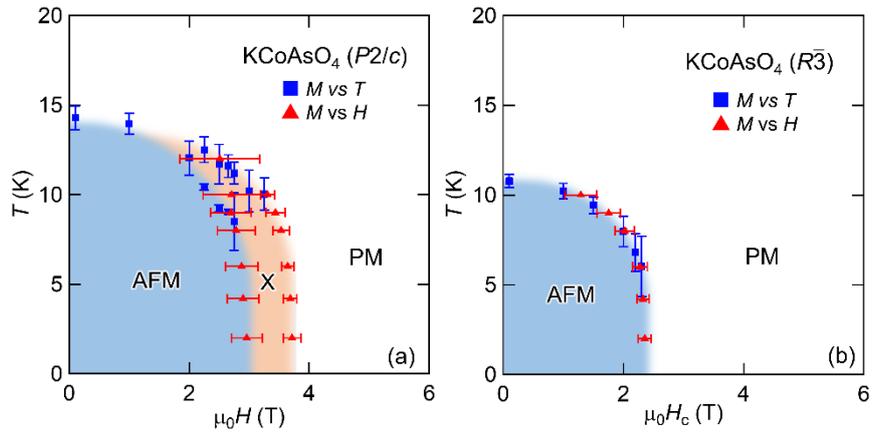

**Fig. 5** Magnetic field versus temperature phase diagram for (a) $P2/c$-type $KCoAsO_4$ and (b) $R\bar{3}$-type $KCoAsO_4$ constructed by magnetic measurement. Squares (blue) and triangles (red) are the transition temperatures and fields determined from the temperature dependence of magnetic susceptibilities ($M$ versus $T$), isothermal magnetization curves ($M$ versus $H$). PM refers to the paramagnetic phase, AFM to the antiferromagnetic phase, and X to the magnetic-field induced phase. The error bars for each plot represent the sigma values obtained from the Gaussian fitting.